\documentclass[11pt,a4paper]{article}
\usepackage[utf8]{inputenc}
\usepackage[english]{babel}
\usepackage{amssymb, amsmath, enumerate}
\usepackage{graphicx}
\usepackage{epstopdf}
\usepackage{subcaption}
\usepackage{authblk}
\usepackage{textcomp}

\newcommand{\qed}{\nobreak \ifvmode \relax \else
      \ifdim\lastskip<1.5em \hskip-\lastskip
      \hskip1.5em plus0em minus0.5em \fi \nobreak
      \vrule height0.75em width0.5em depth0.25em\fi}

\begin{document}
\title{Lattice investment projects support process model   with corruption}
\author[1]{O.A.~Malafeyev\thanks{o.malafeev@spbu.ru}}
\author{S.A.~Nemnyugin\thanks{s.nemnyugin@spbu.ru}}
\affil[1]{Saint-Petersburg State University,  Russia}
\date{}
\maketitle
\begin{abstract}
Lattice investment projects support process model with corruption is formulated and analyzed. The model is based on the Ising lattice model of ferromagnetic but takes deal with the social phenomenon. Set of corruption agents is considered. It is supposed that agents are placed in sites of the lattice. Agents take decision about participation in
corruption activity at discrete moments of time. The decision may lead to profit or to loss. It depends on prehistory of the system. Profit and its dynamics are defined by stochastic Markov process. Stochastic nature of the process models influence of external and individual factors on agents profits. The model is formulated algorithmically and is studied by means of computer simulation. Numerical results are given which demonstrate different asymptotic state of a corruption network for various conditions of simulation.  
\end{abstract}



\section{Introduction}

Stochastic methods are widely used in studies of complex systems which operate in complex environment under influence of many unpredictable factors \cite{Never},\cite{Nem}.  

Both dynamics and equilibrium states of social-economic systems are studied by various methods and approaches including models similar to those which are used to study physical phenomena~\cite{Binder}. Stochastic models and other kinds of models are also used~\cite{Redin},\cite{Boit}. The problem of corruption attracts significant attention of researchers because corruption phenomena may have significant social-economic consequences. Various approaches to the problem are presented in [1--30,40--49].

In the present article the model is proposed to study possible corruption states of a set of agents. These kind of phenomena is caused by social relations of agents and state of the economics and/or public administration. In Section~1 model is formulated. In Section~2 numerical results are given.

\section{Model}

In the model social relations are presented by links which connect nodes representing agents of corruption activity. Length of all links is same. We suppose that
both intensity of relations is equal for all agent-to-agent relations and coordination number which characterizes number of social links of an agent is fixed. Such conditions correspond to the Cartesian lattice. For given topology of the lattice the coordination number defines dimensionality of the lattice. It is supposed also that agents of corruption activity may be in one of $q$ states. 

Social dynamic in the model is supposed to be Markovian. Time is discretized and divided onto equal intervals. Every discrete moment randomly or regularly chosen agent takes decision about change of his state.

In simplest case $q = 2$. Let us suppose that one state corresponds to the decision to participate in corruption activity and second one to non participation in this activity. Transition to the first state corresponds to agent's profit. Second kind of transition leads to his loss. Every decision is taken with accounting of agent interactions with other participants of corruption activity.  Effect of boundary is excluded by imposing periodic boundary conditions.

\begin{figure}[htb]
	\begin{center}
		\includegraphics[scale=0.5]{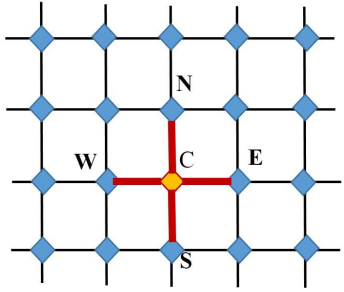}
		\caption{Topology of two-dimensional model and social relations between agents of corruption activity.}\label{fig1}
	\end{center}
\end{figure}

Topology of two-dimensional model and social relations are presented on Fig.~\ref{fig1}. In the article three-dimensional lattice model is studied so Fig.~\ref{fig1} may be considered as two-dimensional section of the models lattice along one of its planes.

Let $c_i$ --- is the state of $i$-th agent and ${\bf c} = \{c_1, c_2, ..., c_M\}$ --- is the state vector of the social system. $M$ --- is number of agents. Dynamic of the system is defined by the function:
$$
W({\bf c}) = \sum_{i = 1}^M \phi_i({\bf c}), 
$$
where functions $\phi_i({\bf c})$ describe interactions between agents. Dynamics is also defined by the procedure of decision making about change of the state of $i$-th agent. Algorithm of modelling of decision making is analogous to Metropolis algorithm, which is widely used for numerical study of lattice systems in physics~\cite{Binder}.

As an example following definition of the function $\phi$ may be given:
$$
\phi_i({\bf c}) = -J \sum_{j = \{S, W, E, N, U, D\}} c_i c_j, 
$$
where $\{S, W, E, N, U, D\}$ --- are nearest neighbours of the $i$-th agent (six for the Cartesian three-dimensional lattice), $J$ --- is constant, which is one of the model's parameters and describes intensity of social interactions between agents. In more complicated models set ${\bf J} = \{J_{11}, J_{12}, ..., J_{MM}\}$ may be used as well as other definition of the function $\phi$:
$$
\phi_i({\bf c}) = - \sum_{j = \{S, W, E, N, U, D\}} J_{ij} c_i c_j. 
$$
This case corresponds to the spin glass model. Various kinds of lattices may be also used. Here $\beta$ --- is the model's parameter. Its inverse value $\beta = \frac{1}{T}$ characterizes social-economic activity. Less values of $T$ correspond to more stagnation of the social-economic activity. 

In according to the Metropolis algorithm initial state of the system may be chosen randomly then at every step an agent ($k$) is taken randomly and his state is replaced by the trial one:
$$
c_k \leftarrow (c_k)_{trial}.
$$
Replacement of the previous state by the trial one leads to the change of the value of the function $W$:
$$
\Delta W({\bf c}) = W({\bf c}_{trial}) - W({\bf c})
$$. 
\noindent If $\Delta W({\bf c}) < 0$, then trial state must be taken as the next state of the Markov chain. If $\Delta W({\bf c}) > 0$, then pseudorandom number $\xi$ is generated which belongs to the interval [0,~1] and if $\exp{(-\beta \Delta W)} \ge \xi$, then trial state is taken as the next state of the Markov chain otherwise the state is not changed.

From the point of view of social dynamics this algorithm may be interpreted as follows. At discrete time moment some agent feels temptation to take part in corruption activity. He tries to estimate possible profit and risks of this decision. Decision is based both on external factors and own life experience which in part is defined by random facts. Metropolis accept and rejection rule is one of possible models of social dynamics. Its reliability should be verified by the statistical data.

Despite of the initial state of the system stochastic evolution governed by Metropolis algorithm converges to the unique final distribution.
Modelling of stochastic dynamics of a system let us estimate influence of characteristics of social interactions and state of economic or public administration on transition processes and setting of the stationary state of corruption activity.

Total profit of agents from taking part in corruption activity may be defined as follows:
$$
U = \sum_{j: c_j > 0} c_j, 
$$
It characterizes loss of economic due to corruption activity. Average size of clusters of similar values of state variables are characteristics of scale of corruption.

\section{Numerical results}

\begin{figure}[htb]
	\begin{center}
		\includegraphics[scale=0.8]{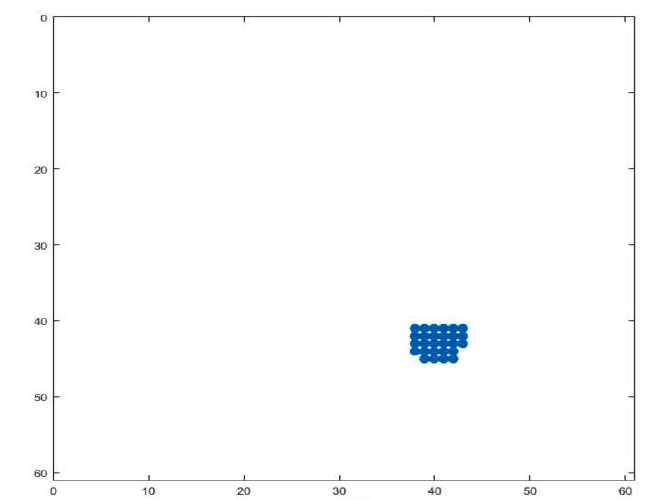}
		\includegraphics[scale=0.8]{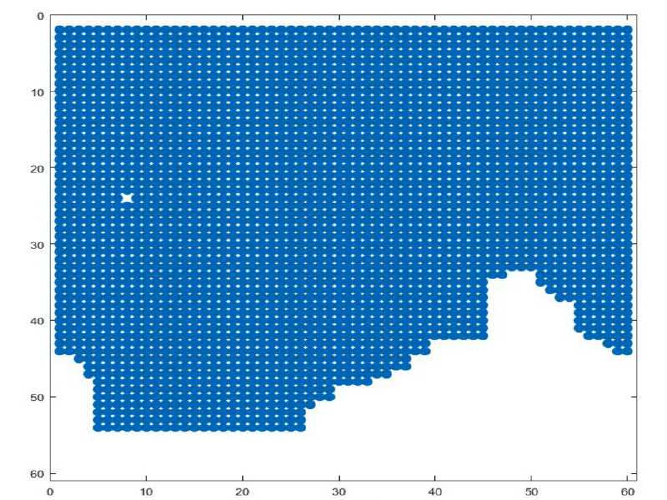}
		\caption{Examples of configurations for the lattice model and the case of low activity (T = 0.5)}\label{fig2}
	\end{center}
\end{figure}

\begin{figure}[htb]
	\begin{center}
		\includegraphics[scale=0.8]{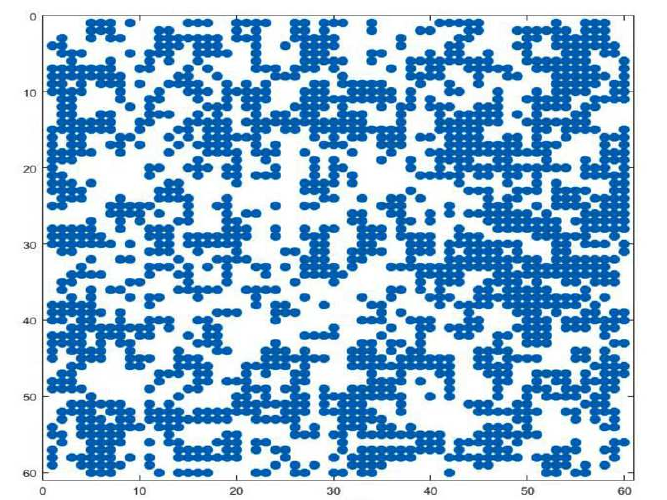}
		\includegraphics[scale=0.8]{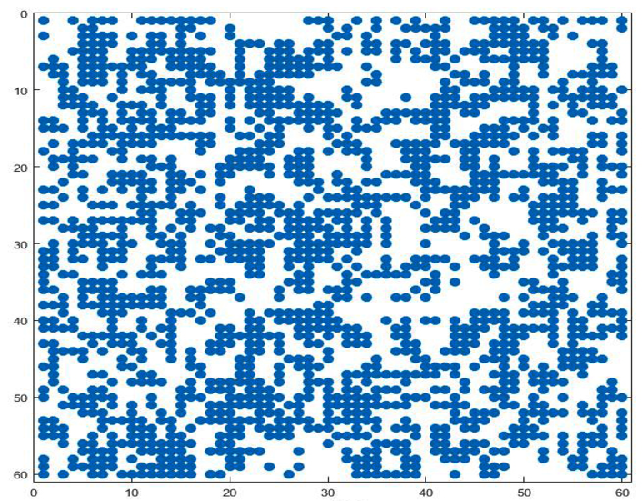}
		\caption{Examples of configurations for the lattice model and the case of high activity (a) T= 4.0; (b) T = 5.0}\label{fig3}
	\end{center}
\end{figure}

Numerical simulations have been performed for the simple model of corruption activity on cartesian three-dimensional lattice with size 60x60x60 and two possible local states with periodic boundary conditions. Number of Markov steps was taken to be $10^9$. This value guarantees for the lattice of a given size that transition to the final stationary distribution is nearly completed. Examples of configurations for low and high economic activity cases are given on Fig.~\ref{fig2} and Fig.~\ref{fig3}) respectively. Figures demonstrate two-dimensional sections of the three-dimensional lattice. It may be seen that in the case of low social activity dense clusters of various sizes are appeared and corruption has tendency to be widespread. 

A cluster is defined as the set of connected agents which made decision to take part in the corruption activity. As a rule it doesn't contains agents of other kind. 

High social-economic activity leads to subdivision of clusters into smaller parts interleaving with groups of agents who don't participate in the corruption activity. Moreover second case show faster dynamics with appearing and disappearing of small clusters.

\section{Conclusion}

In the article simple investment projects support process model with corruption is proposed. It is formulated as three-dimensional lattice model with nearest neighbours interactions and periodic boundary conditions. Due to this model agents of corruption activity take decisions about participation in corruption activity at discrete moments of time. Qualitative results of simulation seems to be reasonable but should be verified by empirical data.

The study was performed using computational resources provided by the Resource Physics Educational Centre of the Research park of Saint-Petersburg State University.

\section{Acknowledgements}
The work is partly supported by work RFBR No. 18-01-00796.

\end{document}